\documentclass[aps,prl,nofootinbib,twocolumn,preprintnumbers,nobalancelastpage,superscriptaddress]{revtex4}

\usepackage[english]{babel}
\usepackage{amsmath,amssymb}
\usepackage[dvips]{graphicx}
\usepackage[sort&compress]{natbib}
\usepackage{amsfonts}
\usepackage{bbm}
\usepackage{color}

\DeclareMathAlphabet{\mathsc}{OT1}{cmr}{m}{sc}

\allowdisplaybreaks

\def\10{$SO(10)$}
\def\21{SU(2) $\otimes$ U(1) }

\def\422{$SU(4) \otimes SU(2) \otimes SU(2)$}
\def\321{SU(3) $\otimes$ SU(2) $\otimes$ U(1)}

\def\lsim{\raise0.3ex\hbox{$\;<$\kern-0.75em\raise-1.1ex\hbox{$\sim\;$}}}
\def\gsim{\raise0.3ex\hbox{$\;>$\kern-0.75em\raise-1.1ex\hbox{$\sim\;$}}}

\newcommand{\flux}[2][]{\ensuremath{\ifthenelse{\equal{#1}{}}{}{^{#1}\!}\mathit{#2}}}

\def \snu  { {\tilde{\nu}}}

\newcommand{\GeV}{\, {\rm GeV}}

\newcommand{\TeV}{\, {\rm TeV}}

\newcommand{\AddrAHEP}{%
  AHEP Group, Instituto de F\'{\i}sica Corpuscular --
  C.S.I.C./Universitat de Val{\`e}ncia \\
  Edificio Institutos de Paterna, Apt 22085, E--46071 Valencia, Spain}
\newcommand{\AddrTo}{%
  Dipartimento di Fisica Teorica,
  Universit\`a di Torino and INFN/Torino \\
  Via Pietro Giuria 1, 10125 Torino, Italy}
\newcommand{\AddrLisb}{%
 Departamento de F\'\i sica and CFTP, Instituto Superior T\'ecnico\\
          Av. Rovisco Pais 1, 1049-001 Lisboa, Portugal }

\begin{document}

\preprint{DFTT 11/2008, IFIC/08-31}

\title{Minimal supergravity sneutrino dark matter and inverse seesaw neutrino
  masses}

\author{C. Arina}\affiliation{\AddrTo}
\author{F. Bazzocchi}\affiliation{\AddrAHEP}
\author{N. Fornengo}\affiliation{\AddrTo}
\author{J. C. Romao}\affiliation{\AddrLisb}
\author{J. W. F. Valle}\affiliation{\AddrAHEP}

\begin{abstract}
  We show that within the inverse seesaw mechanism for generating
  neutrino masses minimal supergravity is more likely to have a
  sneutrino as the lightest superparticle than the conventional
  neutralino.
  We also demonstrate that such schemes naturally reconcile the small
  neutrino masses with the correct relic sneutrino dark matter
  abundance and accessible direct detection rates in nuclear recoil
  experiments.
  \end{abstract}
\maketitle

\section{Introduction}

Over the last fifteen years we have had solid experimental
evidence for neutrino masses and oscillations~\cite{Maltoni:2004ei},
providing the first evidence for physics beyond the Standard Model. 
On the other hand, cosmological studies clearly show that a large
fraction of the mass of the Universe in dark and must be
non--baryonic.

The generation of neutrino masses may provide new insight on the
nature of the dark matter~\cite{Berezinsky:1993fm}.
In this Letter we show that in a minimal supergravity (mSUGRA) scheme
where the smallness of neutrino masses is accounted for within the
inverse seesaw mechanism the lightest supersymmetric particle is
likely to be represented by the corresponding neutrino superpartner
(sneutrino), instead of the lightest neutralino. 
This opens a new window for the mSUGRA scenario. Here we consider the
implications of the model for the dark matter issue.
We demonstrate that such a model naturally reconciles the small
neutrino masses with the correct relic abundance of sneutrino dark
matter and experimentally accessible direct detection rates. 

\section{Minimal SUGRA inverse seesaw model}
\label{Model}

Let us add to the Minimal Supersymmetric Standard Model (MSSM) three
sequential pairs of \21 singlet neutrino superfields ${\widehat
  \nu^c}_i$ and $\widehat{S}_i$ ($i$ is the generation index), with
the following superpotential
terms~\cite{mohapatra:1986bd,Deppisch:2004fa},
\begin{equation} 
{\cal W} =  {\cal W_{\rm MSSM}}  +\varepsilon_{ab}\,
 h_{\nu}^{ij}\widehat L_i^a\widehat \nu^c_j\widehat H_u^b
+ M_{R}^{ij}\widehat \nu^c_i\widehat S_j 
+\frac{1}{2}\mu_S^{ij} \widehat S_i \widehat S_j 
\label{eq:Wsuppot} 
\end{equation}
where ${\cal  W_{\rm MSSM}}$  is the usual MSSM superpotential.
In the limit $\mu_S^{ij} \to 0$ there are exactly conserved lepton
numbers assigned as $(1,-1,1)$~\cite{mohapatra:1986bd,Deppisch:2004fa}
for $\nu$, $\nu^{c}$ and $S$, respectively.

The extra singlet superfields induce new terms in the soft--breaking
Lagrangian:
\begin{eqnarray}
\mathcal{-L}_{\rm soft} &=& \mathcal{-L}_{\rm soft}^{\rm MSSM} + 
\tilde{\nu}^c_i\ \mathbf{M^2_{\nu^c}}_{ij} \tilde{\nu}^c_j + \tilde{S}_i\,
\mathbf{M^2_{S}}_{ij} \tilde{S}_j \\
& & + \varepsilon_{ab}\,
 A_{h_{\nu}}^{ij}  \tilde{L}_i^a  \tilde{\nu}^c_j  H_u^b
+ B_{M_{R}}^{ij}  \tilde{\nu}^c_i \tilde{S}_j 
+\frac{1}{2} B_{\hat{\mu_S}}^{ij}  \tilde{S}_i  \tilde{S}_j \nonumber
\label{eq:soft}
\end{eqnarray}
where $\mathcal{L}_{\rm soft}^{\rm MSSM}$ is the MSSM SUSY--breaking
Lagrangian.

Small neutrino masses are generated through the inverse seesaw
mechanism~\cite{mohapatra:1986bd,Deppisch:2004fa,Nunokawa:2007qh}: the
effective neutrino mass matrix $m^{\rm eff}_{\nu}$ is obtained by the
following relation:
\begin{equation}
  \label{eq:1}
  m^{\rm eff}_{\nu}= -v_u^2 h_{\nu} \left(M_R^T\right)^{-1} {\mu_S}
    M_R^{-1} h_{\nu}^T = \left(U^T\right)^{-1} m_{\mu}^{\rm diag}\ U^{-1}
\end{equation}
where $h_\nu$ defines the Yukawa matrix and $v_u$ is the $H_u$ vacuum
expectation value. The smallness of the neutrino mass is ascribed to
the smallness of the $\mu_S$ parameter, rather than the largeness of
the Majorana--type mass matrix $M_R$, as required in the standard
seesaw mechanism~\cite{Nunokawa:2007qh}. In this way light (eV scale
or smaller) neutrino masses allow for a sizeable magnitude for the
Dirac--type mass $m_D=v_u h_\nu$ and a TeV--scale mass for the
right-handed neutrinos, features which have been shown to produce an
interesting sneutrino dark matter phenomenology~\cite{Arina:2007tm}.

The main feature of our model is that the nature of the dark matter
candidate, its mass and couplings all arise from the same sector
responsible for the generation of neutrino masses.
In order to illustrate the mechanism we consider the simplest
one-generation case, for simplicity. In this case where the sneutrino mass
matrix reads
\begin{eqnarray}
 \mathcal{M}^2  = 
 \begin{pmatrix}
\mathcal{M}^2_+ & \mathbf{0}\cr
 \mathbf{0} & \mathcal{M}^2_-\cr
\end{pmatrix}
\end{eqnarray}
where the two sub--matrices $\mathcal{M_\pm}^2$ are:
\begin{widetext}
\begin{eqnarray}\label{eq:snumatrix}
\mathcal{M_{\pm}}^2  = 
 \begin{pmatrix}
m^2_L+\frac{1}{2} m^2_Z \cos 2\beta+m^2_D & \pm (A_{h_{\nu}}v_u-\mu m_D {\rm cotg} \beta) & m_D M_R\cr
 \pm (A_{h_{\nu}}v_u-\mu m_D {\rm cotg}\beta) & m^2_{\nu^c}+M_R^2+m^2_D & \mu_S M_R \pm B_{M_R}\cr
m_D M_R & \mu_S M_R \pm B_{M_R} & m^2_S+\mu^2_S+M^2_R\pm B_{\mu_S}
\end{pmatrix}
\end{eqnarray}
\end{widetext}
in the CP eigenstates basis: $\Phi^{\dag} = (\snu_{+}^\ast
\,\tilde{\nu}_{+}^{c\ast} \, \tilde{S}_+^\ast \,\, \snu_-^\ast \,
\tilde{\nu}_-^{c\ast} \, \tilde{S}_-^\ast)$.
Once diagonalized, the lightest of the six mass eigenstates is our
dark matter candidate and it is stable by $R$--parity conservation.

\section{A novel supersymmetric spectrum}

\begin{figure}[t!]
\hspace{-1cm}
\includegraphics[width=\columnwidth]{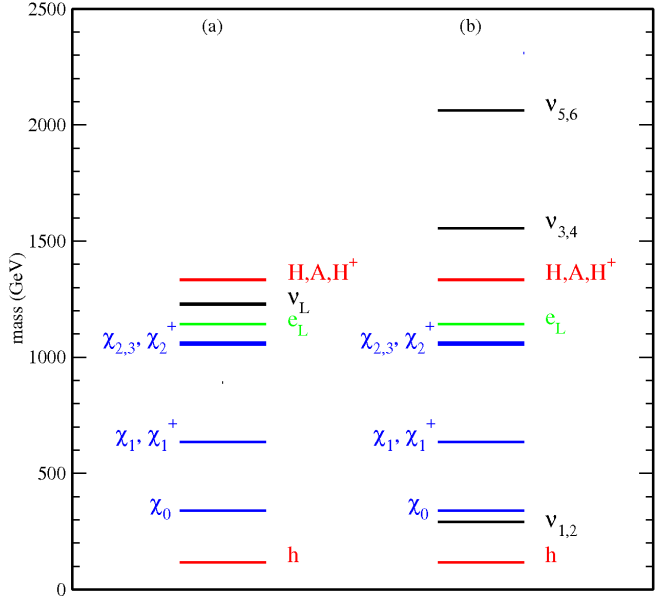}
\caption{Supersymmetric particle spectrum in the standard MSUGRA
  scheme [panel (a)] and in the inverse seesaw mSUGRA model [panel (b)]
  with parameters chosen as: $m_0= 358$ GeV, $m_{1/2}= 692$ GeV, $A_0
  = 0$, $\tan\beta=35$ and sign $\mu >0$. The sneutrino sector has the
  additional parameter $B_{\mu_S}$, fixed at 10 GeV$^2$. The squark
  sector is not shown. }
\label{fig:spectrum}
\end{figure}
\begin{figure}[t!]
\hspace{-1cm}
\includegraphics[width=\columnwidth]{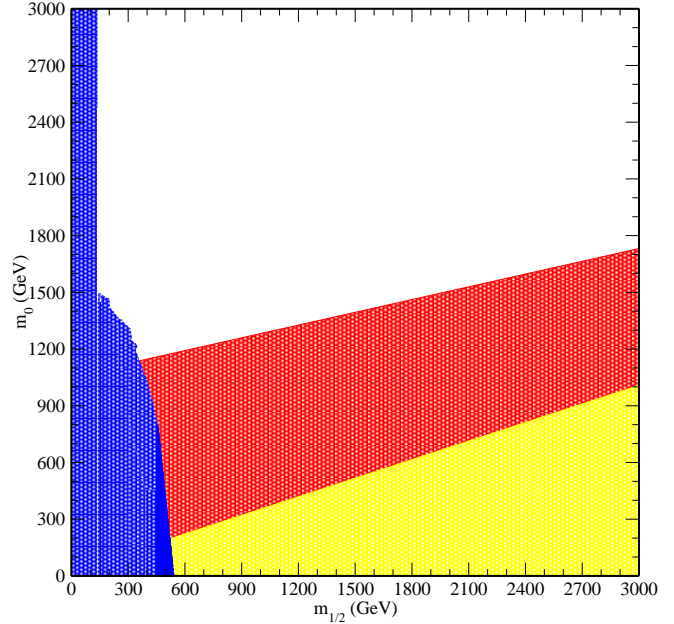}
\caption{The $m_{0}-m_{1/2}$ plane for $\tan\beta=35$, $A_0=0$ and
  $\mu>0$.  
The red and yellow areas denote the set of
  supersymmetric parameters where the sneutrino is the LSP in inverse
  seesaw models (notice that it includes all the yellow region where the
  $\tilde{\tau}$ is the LSP in the standard mSUGRA case). 
The white
  region has the neutralino as LSP in both standard and modified
  mSUGRA.  For the sneutrino LSP region, the additional parameters are: 
$B_{\mu_S}= 10 \GeV^2$, $M_R=500 \GeV$, $m_D=50 \GeV$ and $\mu_S=1$ eV.
The blue region is excluded (see text). 
}
\label{fig:m0mh12}
\end{figure}

Let us now consider the model within a minimal SUGRA scenario. In the
absence of the singlet neutrino superfields, the mSUGRA framework
predicts the lightest supersymmetric particle (LSP) to be either a
stau or a neutralino, and only the latter case represents a viable
dark matter candidate. In most of the mSUGRA parameter space, however,
the neutralino relic abundance turns out to exceed the WMAP bound
\cite{Komatsu:2008hk} and hence the cosmologically acceptable regions
of parameter space are quite restricted.

In contrast, when the singlet neutrino superfields are added, a
combination of sneutrinos emerges quite naturally as the LSP.  Indeed,
we have computed the resulting supersymmetric particle spectrum and
couplings by adapting the SPheno code \cite{Porod:2003um} so as to
include the additional singlet superfields. An illustrative example of
how the minimal supergravity particle spectrum is modified by the
presence of such states is given in Fig. \ref{fig:spectrum}. This
figure shows explicitly how a sneutrino LSP is in fact realized.

A more general analysis in the mSUGRA parameter space is shown in
Fig. \ref{fig:m0mh12}: the dark (blue) shaded area is excluded either
by experimental bounds on supersymmetry and Higgs boson searches, or
because it does not lead to electroweak symmetry breaking, while the 
(light) yellow region refers to stau LSP in the conventional
(unextended) mSUGRA case.
As expected, in all of the remaining region of the plane, the
neutralino is the LSP in the standard mSUGRA case. The new
phenomenological possibility which opens up thanks to the presence of
the singlet neutrino superfields where the sneutrino is the LSP
corresponds to the full mid-gray (red) and light (yellow) areas. In
what follows we demonstrate that in this region of parameter space
such a sneutrino reproduces the right amount of dark matter and is not
excluded by direct detection experiments.

\section{Sneutrino LSP as Dark Matter}
\begin{figure}[t]
\hspace{-1cm}
\includegraphics[width=\columnwidth]{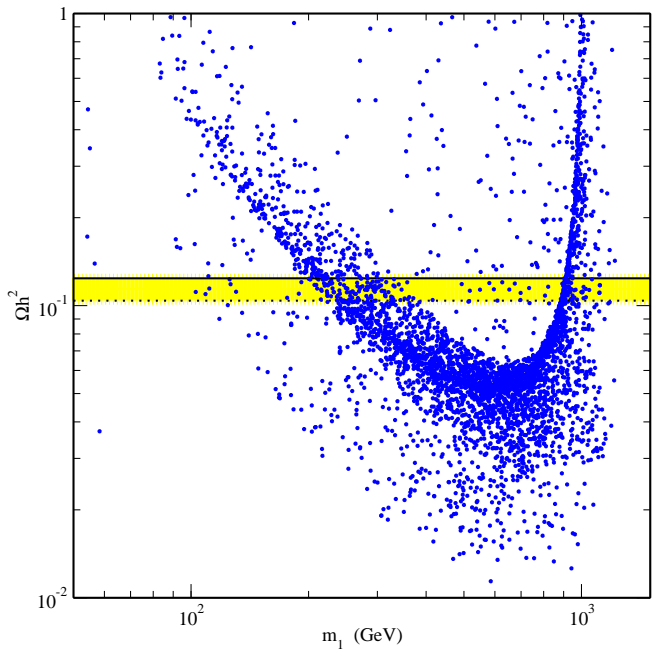}
\caption{Sneutrino relic abundance $\Omega h^2$ as a function of the
  LSP sneutrino mass $m_1$, for a full scan of the supersymmetric
  parameter space: $100 \GeV < m_0 < 3 \TeV$, $100 \GeV < m_{1/2}< 3
  \TeV$, $1 \GeV^2 <B_\mu < 100 \GeV^2$, $A_0=0$, $3 < \tan\beta<50$,
  $10^{-9} \GeV <\mu_S<10^{-6} \GeV$. The yellow band delimits the
  WMAP~\cite{Komatsu:2008hk} cold dark matter interval at 3 $\sigma$
  of C.L.: $0.104 \leq \Omega_{\rm{CDM}} h^2 \leq 0.124$.}
\label{fig:omega}
\end{figure}
\begin{figure}[t]
\hspace{-1cm}
\includegraphics[width=\columnwidth]{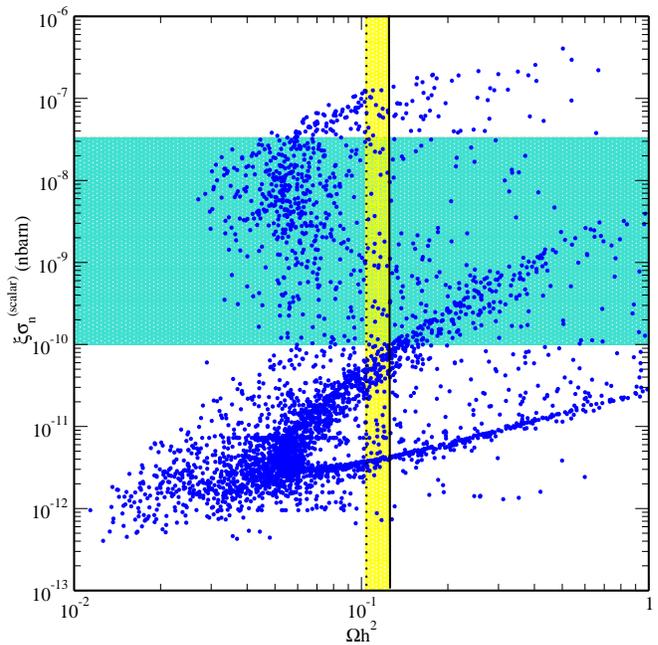}
\caption{Sneutrino--nucleon scattering cross section $\xi \sigma^{\rm
    (scalar)}_{\rm nucleon}$ vs. the sneutrino relic abundance $\Omega
  h^{2}$, for the same scan of the supersymmetric parameter space
  given in Fig.~\ref{fig:omega}. The horizontal [light blue] band
  denotes the current sensitivity of direct detection experiments; the
  vertical [yellow] band delimits the 3 $\sigma$ C.L. WMAP 
   cold dark matter range~\cite{Komatsu:2008hk}.}
\label{fig:direct}
\end{figure}

The novelty of the spectrum implied by mSUGRA implemented with the
inverse seesaw mechanism is that it may lead to a bosonic dark matter
candidate, the lightest sneutrino $\tilde{\nu}_1$, instead of the
fermionic neutralino. To understand the physics it suffices for us to
consider the simple one sneutrino generation case~\footnote{We adopt
  the same approximation used in the relic density calculation within
  the standard minimal mSUGRA model, which we have checked holds in
  our case as well.}. The relic density of the sneutrino candidate is
shown in Fig. \ref{fig:omega}.  The lightest mass eigenstate is also a
CP eigenstate and coannihilates with the NLSP, a corresponding heavier
opposite--CP sneutrino eigenstate. We notice that this situation
provides a nice realization of inelastic dark matter, a case where the
dark matter possesses a suppressed scattering with the nucleon,
relevant for the direct detection scattering cross section, shown in
Fig. \ref{fig:direct}.

From Fig. \ref{fig:omega} we see that a large fraction of the
sneutrino configuration are compatible with the WMAP cold dark matter
range, and therefore represents viable sneutrino dark matter
models. Fig. \ref{fig:direct} in addition shows that direct detection
experiments do not exclude this possibility: instead, a large fraction
of configurations are actually compatible and under exploration by
current direct dark matter detection experiments. This fact is partly
possible because of the inelasticity characteristics we have mentioned
above, which reduces the direct detection cross section to acceptable
levels \cite{Arina:2007tm}.

We stress that all models reported in Figs. \ref{fig:omega} and
\ref{fig:direct} have the inverse seesaw-induced neutrino masses
consistent with current experimental observations for natural values
of its relevant parameters. We also note that the lepton--number
violating parameter $B_{\mu_S}$, which determines the lightest mass
sneutrino eigenstate and its couplings, also has an impact on the
neutrino sector, since it can induce one-loop corrections to the
neutrino mass itself (for details, see Ref. \cite{Arina:2007tm} and
references therein). These corrections must be small, in order not to
go into conflict with the bounds on neutrino masses, and this in turn
implies that the mass splitting between the sneutrino LSP and
sneutrino NLSP is small (less than MeV or so)~\cite{Arina:2007tm},
implying the inelasticity of the sneutrino scattering with nuclei
\cite{Arina:2007tm}.  The parameter $\mu_s$ therefore plays a key role
in controlling the neutrino mass generation, the sneutrino relic
abundance and the direct detection cross section.

In conclusion, in this Letter we have presented an mSUGRA scenario in
which neutrino masses and dark matter arise from the same sector of
the theory. Over large portions of the parameter space the model
successfully accommodates light neutrino masses and sneutrinos dark
matter with the correct relic abundance indicated by WMAP as well as
direct detection rates searches consistent with current dark matter
searches.
The neutrino mass is generated by means of an inverse
seesaw mechanism, while in a large region of parameters the dark
matter is represented by sneutrinos. The small superpotential mass
parameter $\mu_S$ controls most of the successfull phenomenology of
both the neutrino and sneutrino sector. 
In the absence of $\mu_S$ neutrinos become massless, Eq.~(\ref{eq:1}).
The bilinear superpotential term $\mu_S^{ij} \widehat S_i \widehat
S_j$ could arise in a spontaneous way in a scheme with an additional
lepton-number-carrying singlet superfield $\sigma$, implying the
existence of a majoron~\cite{gonzalez-garcia:1988rw}. In this case,
the dominant decays of the Higgs boson are likely to be into a pair of
majorons~\cite{Joshipura:1992hp}. Such invisible mode would be
``seen'' experimentally as missing momentum, but the corresponding
signal did not show up in the LEP
data~\cite{Abdallah:2003ry}. Although hard to catch at the LHC such
decays would provide a clean signal in a future ILC facility.
Similarly, the standard bilinear superpotential term $\mu H_u H_d$
present in the minimal supergravity model could also be substituted by
a trilinear, in a NMSSM-like scheme~\cite{CerdenoMunoz:2008}. 

Note that our proposed scheme may also have important implications for
supersymmetric particle searches at the LHC, due to modified particle
spectra and decay chains.
Additional experimental signatures could be associated with the
(quasi-Dirac) neutral heavy leptons formed by $\nu^c$ and $S$, whose
couplings and masses are already restricted by LEP
searches~\cite{Dittmar:1989yg,Abreu:1996pa}.

{\sl Acknowledgements.} We warmly thank M. Hirsch for stimulating discussions.
This work was supported by MEC grant
FPA2005-01269, by EC Contracts RTN network MRTN-CT-2004-503369 and
ILIAS/N6 RII3-CT-2004-506222, by FCT grant POCI/FP/81919/2007 
and by research grants funded jointly by the Italian Ministry of Research
 and by the Istituto Nazionale di Fisica Nucleare
(INFN) within the {\sl Astroparticle Physics Project}.

\bibliographystyle{h-physrev4}

\end{document}